# Searching for behavioral homologies:
# Shared generative rules for expansion and narrowing down of the locomotor repertoire in Arthropods and Vertebrates

Alex Gomez-Marin[a1], Efrat Oron[b], Anna Gakamsky[b],
Dan Valente[c], Yoav Benjamini[d] and Ilan Golani[b1]

[a]Champalimaud Neuroscience Programme, Lisbon, Portugal; [b]Department of Zoology, Faculty of Life Sciences and Sagol School of Neuroscience, Tel Aviv University, Israel; [c]Cold Spring Harbor laboratory, USA; [d]Department of Statistics and Sagol School of Neuroscience, Tel Aviv University.

[1]To whom correspondence should be addressed: agomezmarin@gmail.com, ilan99@tau.ac.il

(Running title: **A common mobility gradient across phyla**)

Abstract

We use immobility as an origin and reference for the measurement of locomotor behavior; speed, the direction of walking and the direction of facing as the three degrees of freedom shaping fly locomotor behavior, and cocaine as the parameter inducing a progressive transition in and out of immobility. In this way we expose and quantify the generative rules that shape fruit fly locomotor behavior, which consist of a gradual narrowing down of the fly's locomotor freedom of movement during the transition into immobility and a precisely opposite expansion of freedom during the transition from immobility to normal behavior. The same generative rules of narrowing down and expansion apply to vertebrate behavior in a variety of contexts, Recent claims for deep homology between the vertebrate basal ganglia and the arthropod central complex, and neurochemical processes explaining the expansion of locomotor behavior in vertebrates could guide the search for equivalent neurochemical processes that mediate locomotor narrowing down and expansion in arthropods.

We argue that a methodology for isolating relevant measures and quantifying generative rules having a potential for discovering candidate behavioral homologies is already available and we specify some of its essential features.

**Significance Statement**

Formulating the generative rules that shape behavior across species and phyla is an essential step in the establishment of a comparative study of behavior. Invariant generative rules provide an architectural plan *(bauplan)* that guides research and defines and validates the intrinsic constituents of that *bauplan*. Examining whether the same rules shape arthropod and vertebrate locomotor repertoire requires solutions to questions such as whether a homology should be based first and foremost on structure or on function, on generative rules or on common descent or on both, and how can the invariant core be distinguished from the adaptive envelope of behavior. The significance of this study lies in the new solutions it offers to these old questions: defining an intrinsic origin and measuring the transition from simple to complex and from complex to simple behavior; using actively managed kinematic quantities; suspending judgment about common descent; using the generative rules as a heuristic search image for the discovery of equivalent behaviors and the neural processes that mediate them; and only then offering the established behavioral *bauplan* as candidate for the status of a Darwinian homology, based on demonstrated common descent.

**Introduction**
The establishment of homologies is an indispensable goal in evolutionary biology. In pre-Darwinian comparative anatomy, a homologue has been defined as "The same organ in different animals under every variety of form and function"[1]. Based on this definition, anatomists compared skeletons using validated distinctions like a forelimb, a humerus, a radius and an ulna, and compared brains using validated structures such as thalamus, cortex and striatum. These structures gained their identity and validity as homologues by demonstrating that they occupied the same relative position, and had the same connectivity across a wide array of taxonomic groups[2] and structures sharing the same name and the same morphogenetic history. The validity of these structures has been indispensable for establishing a rigorous science of anatomy. To the same extent, the comparative study of behavior requires the identification of distinct elementary processes, much like skeletal segments and neural structures, which would be established across a wide variety of taxonomic groups on the basis of their connectivity[3], and their moment-to-moment generative history.

In the present study we analyze the morphogenesis of cocaine-induced fruit fly behavior, implementing a strategy and tools that aim at exposing generative rules having the potential of becoming universal right from the very first description of a newly studied behavior.

The accumulation of detailed descriptions of arthropod and vertebrate movement makes the issue of shared principles of organization in the behavior of these taxonomic groups increasingly accessible for comparison. An opportunity for such comparison is offered by the report that, when treated with the dopamine reuptake inhibitor cocaine, *Drosophila melanogaster* performs a sequence of stereotyped behavior patterns leading in and out of immobility including locomotion and circling, apparently similar to the sequence observed in rodents[4,5]. This led the researchers who discovered the phenomenon to suggest that the behavior was homologous in the two phyla. That same behavior has been, however, portrayed as aberrant, unusual, and uncontrolled[6] by other researchers, who used a description that highlighted impairment in the functionality of the behavior. Here we analyze the structure of the very same behavior both in its own right, but also using a strategy and tools that aim at obtaining a description that will entice a cross phyletic comparison. Using our structural analysis we derive from the fly's seemingly aberrant behavior the generative rules that shape a substantial component of both arthropod and vertebrate locomotor behavior. Following comparative anatomy we term the set of invariant relations or generative rules that we look for, the architectural plan, or *bauplan*[7-10] of the behavior.

Given that a rigorous comparative study of behavior requires the establishment of behavioral homologies, how is it that cross phyletic behavioral homologies were hardly documented? In this study we suggest that the seed for obtaining a cross phyletic perspective (or for missing it) is sown in the initial measurement phase. The choices that are made at that stage reflect age old controversies on whether to i) define homologies on the basis of generative rules[11,12], common descent[13,14], or both[15], ii) measure the function or the structure of behavior[16], iii) focus on the behavioral level first or mix behavioral, neural and genetic levels from the start[15], iv) use any reference frame or use intrinsic frames of reference for measuring the behavior[17], v) suspend or even ignore judgement about common descent[17,18], and vi) more generally distinguish between the invariant skeleton of a behavior and its adaptive envelope[19]. All these choices are part and parcel of the process of observation, and are unavoidable; they are made, explicitly or implicitly, by each and every observer of behavior, determining from the start the potential for the universality of the obtained description. Since they are inescapable, they might as well be made deliberately, on the basis of one's aims. Our aim, of establishing a rigorous comparative study of behavior, requires the discovery of the behavioral *bauplan* that will have the potential of becoming a *bona fide* behavioral homology. The choices we made are derived from this aim (implying that other aims might justify other choices).

One strategy in deciphering the organization of anatomical structure is following the process of its morphogenetic differentiation from inception to full blown form. The transition from simple to complex provides a view that is not available by studying the final product. We therefore focus on studying fly behavior in a situation involving differentiation from simple to complex, and decay from complex to simple, teasing apart in this way the particulate processes that add on top of each other to compose full blown behavior[17,20] (or are eliminated one after the other to full decay).

While any selection of variables might be useful (informative), only a selection of the key variables that are actually managed actively by the fly has the potential of defining a behavioral *bauplan* that will subsequently prove to be cross phyletic. Conversely, variables or kinematic quantities that prove universal across phyla are more likely to represent perceptual quantities that are actually managed by the brain. A judicious selection of the key variables that describe the behavior is therefore necessary for discovering cross phyletic homologies.

Recent claims for deep homology between the arthropod central complex and the vertebrate basal ganglia[21] provide an opportunity to examine whether the *bauplan* we discovered can be supplemented with a historical perspective. If cocaine induced behavior, which is mediated by these centers, is the same in the two phyla, then the claim for a behavioral homology would be supported by both a generative claim for a common *bauplan* and a historical claim for common descent. Moreover, the shared generative rules of the behavior can provide a specification of the demand[17,22] on the neural activity and network connectivity within and between substructures of the central complex and the basal ganglia. For example, the expansion of a vertebrate's locomotor repertoire, which has been recently attributed to dopaminergic feedforward loops operating in the basal ganglia[23,24], can guide a study of the relations between the arthropod transition out of immobility and the arthropod's central complex.

## Results

***Behavior in and out of immobility: narrowing down of the path's spatial spread and its build up to normal behavior.*** Figure 1 presents the path traced by the center of mass of a single fly walking in the experimental arena throughout the whole 90 minute session. Upon cocaine administration, a complex dynamics of fly movement leads to immobility (marked by the red dot) which is followed by full recovery of movement. The path segment leading to immobility is colored in blue, and the path leading out of it is colored in green. The path traced in **Figure 1A** unfolds in time in **Figure 1B**, highlighting immobility as the origin to which motion converges and from which motion unfolds. As shown in **Figure 1C**, the fly first traces relatively straight paths, which become increasingly more curved culminating with immobility. In **Figure 1D**, transition out of immobility starts with highly curved paths involving many very small circles followed by increasingly straighter paths. The progressive narrowing down of the locomotor path into immobility, and the progressive buildup of the path into normal behavior is quantified for all flies in Figures **1**E and F.

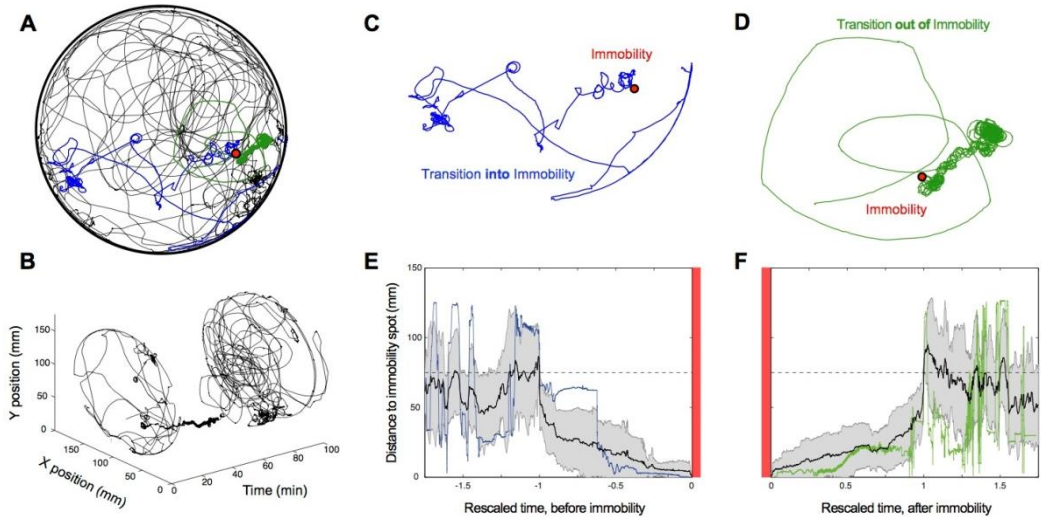

**Figure 1. The locomotor path of cocaine treated flies narrows down into immobility and builds up to spread-out normal behavior.** (**A**) Fly path during the whole 90 minutes session in a circular arena. Red dot indicates location of immobility. Blue colored path depicts transition in and green colored path transition out of immobility. (**B**) Same path as in (A) unfolded in time. (**C**) The transition into immobility is marked by the performance of straight, then increasingly more curved paths, narrowing down the animal's locomotor repertoire. (**D**) The transition out of immobility is marked by the performance of curved, then increasingly straighter paths, widening up the animal's locomotor repertoire. (**E**) By rescaling time in reference to immobility (marked by the vertical red lines), we demonstrate the narrowing of the locomotor path for all flies during transition into immobility, and (**F**) the expansion of the locomotor path for all flies during transition out of immobility.

***Behavior in and out of immobility: narrowing down of the fly's locomotor repertoire and its build up to normal behavior.*** *Defining the three degrees of freedom of locomotor behavior: speed, direction of walking, and body orientation*. The trajectory traced by an animal on a substrate can be characterized by the x-y coordinates as a function of time (**Figure 2A**), from which the velocity vector is calculated. The absolute value of the velocity vector is the speed of the animal (**Figure 2B**). In addition to speed, we can calculate the path curvature (**Figure 2C**). Taking into account the orientation of the longitudinal axis of the fly's body, we can access a third degree of freedom (**Figure 2D**), distinguishing and quantifying in what direction the animal is moving, how fast it is progressing and in what direction it is facing. The left column of **Figure 2** corresponds to these degrees of freedom evolving in time for a short trajectory segment, while the right column illustrates the same degrees of freedom in space. The centroid path is presented in **Figure 2E.** It is color coded according to its curvature in **Figure 2F**. Disk size is proportional to instantaneous speed in **Figure 2G**. Arrows depict the facing direction of the animal in **Figure 2H**. In the illustrated trajectory segment, a burst in speed as the animal traces a relatively straight path is followed by low speed at very high curvature, while the animal rotates in place, at approximately 360 degrees per

second.

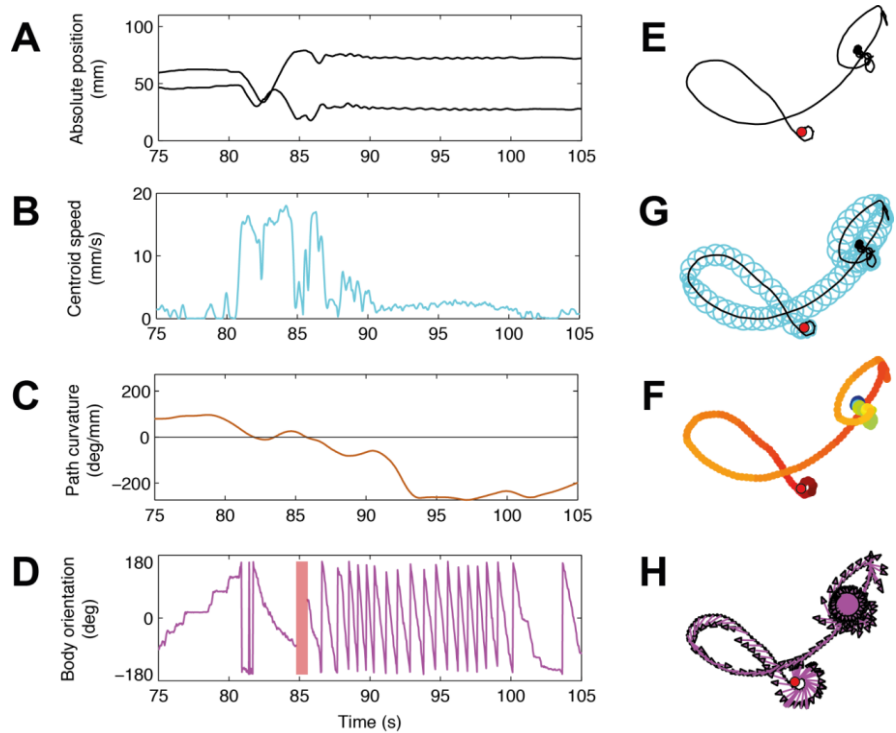

**Figure 2. The three degrees of freedom exercised by a fly at the trajectory level**. Time evolution of XY position (**A**), centroid speed (**B**), path curvature (**C**), and body orientation (**D**), during a 30 second time segment. Same degrees of freedom presented in space: black line represents fly path (**E**), circle diameter depicts speed (**G**), color intensity depicts curvature (**F**), and arrows depict body orientation (**H**). Small red dot marks the beginning of the trajectory. Increase in speed is followed by very low speed at high curvature while the fly vigorously rotates in place.

**Figure 3** shows the three degrees of freedom for the whole dynamics of transition into and recovery from immobility: speed (**Figure 3A**), curvature (**Figure 3B**) and rotation (**Figure 3C**). Using extended immobility as a reference we trace the behavior that precedes it, starting with the inflow of cocaine into the arena, and ending with full recovery of the fly marked by the absence of high rotation in place (high changes in body orientation at low translational speed) and the performance of straight paths (high speed at low curvature). As shown, the fly was totally immobile for 10 minutes (grey shaded area). Recovery from immobility starts with very fast whole-body rotations in place. Each diagonal line in magenta stands for a full 360 degrees rotation. The fly performs approximately 50 full rotations in 10 minutes with almost zero translational velocity and extremely high path curvature. The high frequency of rotations gradually decreases, and so does path curvature. At that time, the animal resumes normal forward progression involving relatively straight paths and high velocity. Transition into immobility starts with normal locomotion marked by high speed, low curvature and absence of extensive body rotations. Next, we observe bursts of high velocity followed by medium and then high curvature, which concur with the setting in of rotations at very low translation speeds, culminating with immobility. We can summarize the moment-to-moment dynamics of this fly as the following sequence: predominance of translation, then high curvature, and finally extensive rotation in place, ceasing in immobility, from which the same sequence is performed in reverse. In other words, forward translation is eliminated from the fly's repertoire first, and rotation last, before the onset of immobility (shutdown; [movie S1](movie S1)) rotation is added to the repertoire first, and forward translation last, in the transition out of immobility (warmup; [movie S2](movie S2)). A representation of the progressive narrowing down of degrees of freedom for movement is provided in [movie S3](movie S3), and of progressive expansion in [movie S4](movie S4).

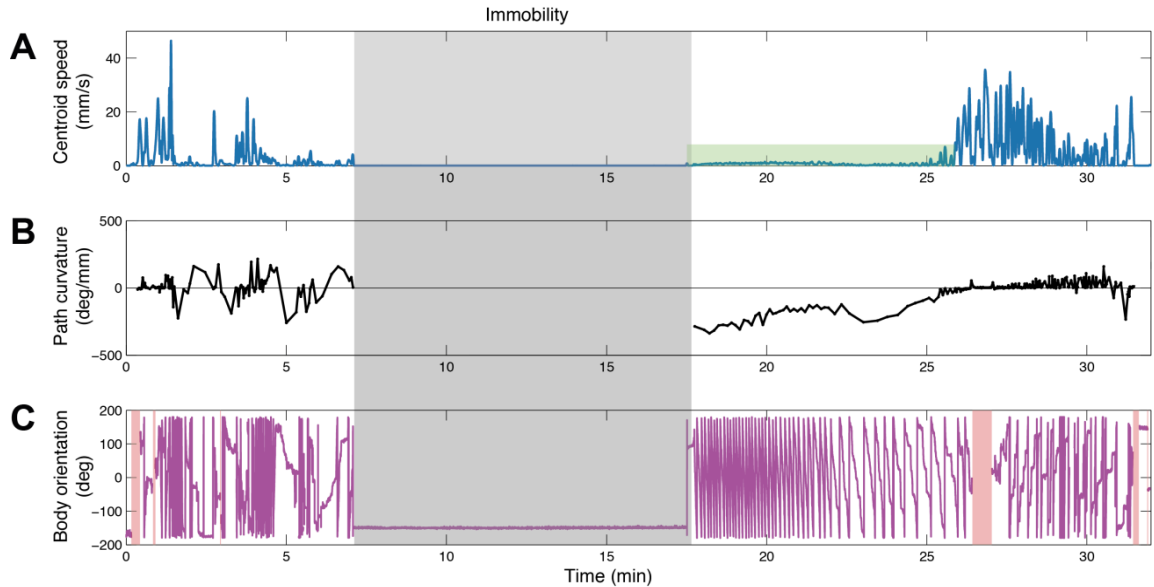

**Figure 3. Moment-to-moment dynamics of the three kinematic degrees of freedom of a fly during a whole session.** The shaded area marks the period of immobility which is used as a reference for the events that precede and follow it. (**A**) The session starts with bursts of speed that progressively decrease towards zero and, after a 10 minutes period of complete immobility, very low speed is followed by normal speed. The green shaded area highlights small but non-zero velocity components at high-curvature during rotation in place, to be studied in depth in Figure 4. (**B**) Straight path is followed by bursts of high curvature until immobility, from which the fly resumes its movement with very high curvature (of the order of a 360 degrees turn in a millimeter) monotonically decreasing to straight paths again. (**C**) Extensive body rotations precede and follow immobility, proceeding from low to high frequency and from high to low frequency. Red shaded areas represent time segments when the fly touches the walls of the arena and body orientation is not tracked. Each diagonal line represents a 360 degrees body rotation. All in all, the session starts with extensive translation, then increasing curvature concurring with frequent body rotations leading into immobility. After immobility, extensive rotation in place concurring with very high path curvature is followed by forward progression along straight paths.

***Behavior out of immobility: walking on highly curved paths at low speed during fast rotation in place near immobility.*** A closer look at path dynamics (**Figure 4A)** reveals that high curvature emerges from small, fast and alternating oscillations in orthogonal components of the velocity vector (**Figure 4B**), that betray the minute circles traced by the animal as it is rotating in place. In particular, during the transition out of immobility, curvature shows a globally monotonical decrease from extremely high values to practically zero curvature. This illustrates that the three independent kinematic degrees of freedom are coordinated. To what extent are they coupled? Mathematically,

speed is independent of curvature and of body orientation, and body orientation is independent of path curvature (see **Methods**). Physically, they can be partly constrained in their relative magnitudes; for example, there is a limit to how fast an organism can go given a certain path curvature, and it is difficult to proceed along a straight line while rotating at high frequency. Biologically, it is an empirical relevant question whether and how these three degrees of freedom are actively managed by the animal. Flies show a range of different velocity components while walking at higher speeds (**Figure 4C**) implying at least partial independence. The freedom in the velocity components (thus, speed and curvature degrees of freedom) is constrained to synchronized oscillations at typical speeds of 1mm/s (**Figure 4D**) which correspond to the fly's body rotating in place.

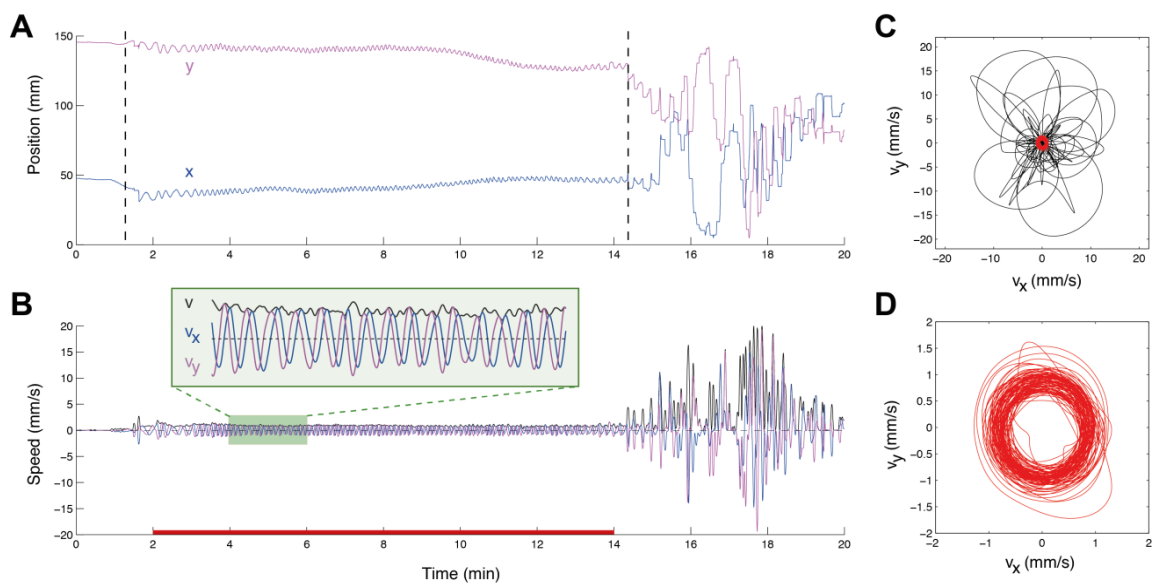

**Figure 4. Detailed dynamics of the transition out of immobility based on velocity components, which determine the dynamics and coordination of speed and curvature**. (**A**) Transition out of immobility is illustrated by plotting the x and y positions as a function of time. Starting from absolute immobility, the fly performs tiny but fast oscillations in the x and y positions, reflecting fast rotations in place, which progressively slow down, and finally change to large displacements, corresponding to normal progression. (**B**) The early stage of transition out of immobility is characterized by very small speeds, whose perpendicular components (vx and vy) alternate in anti-phase oscillatory dynamics, corresponding to very high curvature. Speed along the x and y directions shows the coordinated circling as the fly gets out of immobility. Total speed (v) does not capture the subtleties of rotation in place. (**C**) Phase-plot of speeds along x and y directions, containing low and also high speed progression segments in all directions in a later stage of transition out of immobility. (**D**) Zoom in of the plot in (C) showing only the velocity components during the time interval from minute 2 to minute 14.

***Behavior in and out of immobility: the direction a fly walks and the direction it faces alternate in who leads and who follows***. The relationship between walking direction and facing direction is versatile in flies. For example, flies can walk in any direction while keeping their body orientation fixed (**Figure 5C**), or else facing every which way while proceeding in a specific direction. In intact

flies, the direction of progression changes first, and the direction the fly faces then converges to the new direction set by progression, facing lagging behind by a small angular interval that is quickly closed. The same order of leading and following is exhibited with cocaine, but facing direction lags behind by a much larger angular interval taking a much longer time to close the interval[25]. Here we show that during the stage of transition out of immobility, as the fly rotates along highly curved paths, the two directions of progression and of facing tend to converge to the same values (**Figure 5A-C**). The further away from immobility, the less tight the coupling between these two degrees of freedom progressively gets (**Figure 5D**). Furthermore, the direction of progression leads first while body orientation follows and conversely, in later occasions as the fly regains its freedom of movement away from immobility, body orientation leads and direction of progression follows (**Figure 5E**). In other words, the angular interval between the direction of progression and body orientation is actively managed in two opposite ways, and modulated dynamically with respect to immobility.

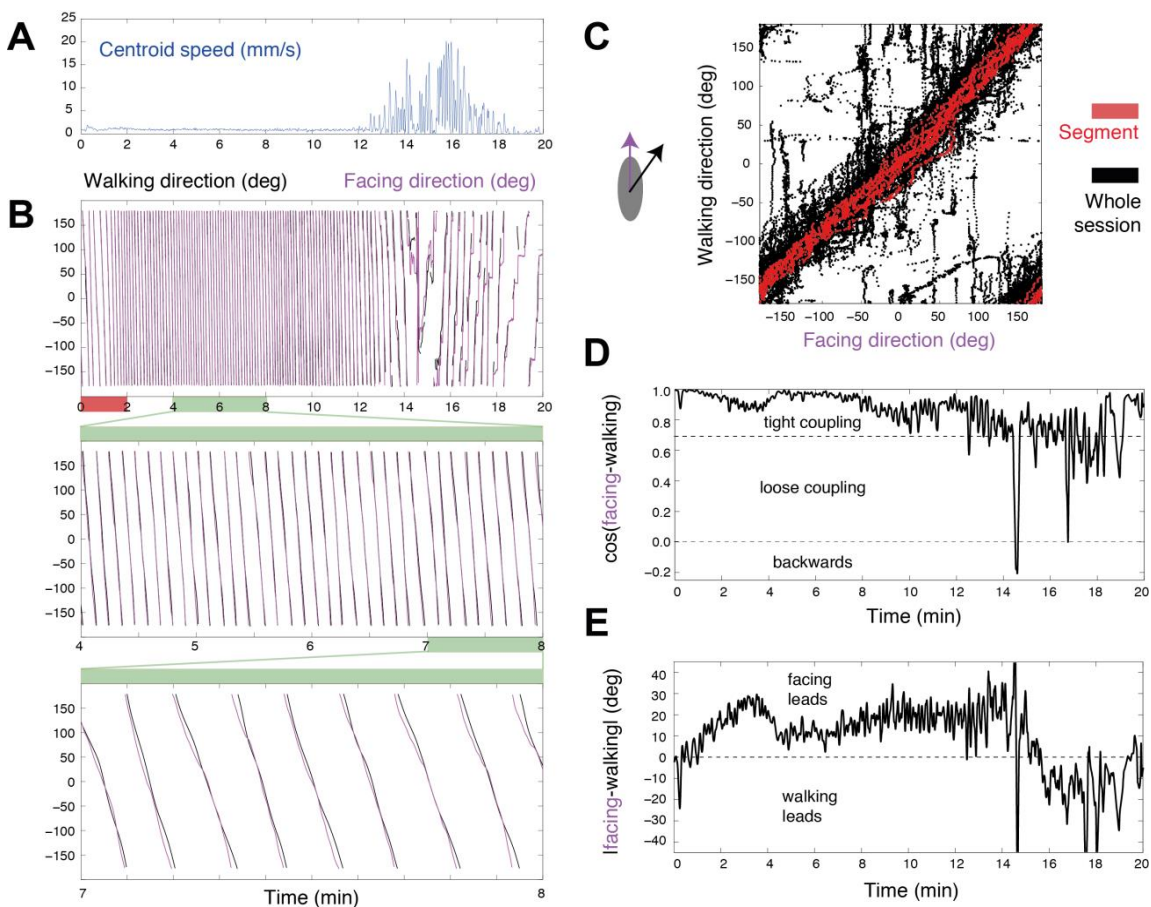

**Figure 5. Active management of facing and walking directions as reflected in the dynamics of curvature and rotation.** (**A**) Centroid speed time course as the fly transitions out of immobility. (**B**) Walking and facing directions during intense rotation in place and high-curvature dynamics. A two minutes segment of the top plot is amplified in the middle plot; and a one minute segment of the middle plot is amplified in the bottom plot, where the tight, but not perfect, coordination between both angles (facing and

walking directions) can be appreciated. (**C**) Phase-plot of facing-walking direction value combinations across a whole session (black), and selected segment showing that close to immobility the coordination is tight (red). (**D**) Cosine of the difference of facing and walking angles, progressively decreasing from 1 (zero difference), to 0.7 (45 degree difference), and reaching below 0 (more than 90 degree difference), quantifying the transition from tighter to looser coupling of rotational and curvature degrees of freedom as the fly gets out of immobility. (**E**) Quantification of the difference in angle between the degree of freedom that leads and the one that follows. Facing leads close to immobility, with walking direction lagging behind up to 30 degrees. Walking direction leads at later stages.

***Behavior in and out of immobility: the generative rules common to all fly sessions.*** Having discussed the relationship between curvature and rotation, we can now concentrate on the relationship between translation and rotation. Since the timescale we examine comprises the whole session dynamics (which can last for more than an hour), we calculate next the cumulative translation by integrating velocity in time and cumulative rotation by unwrapping the body angle (see **Figure 6A**). After smoothing the cumulative measures (see **Methods**) we calculate their derivative, and in this way we obtain the changes in rotation and in translation for the whole session (see **Figure 6B**). Again we use immobility as a reference and measure the global peaks of activity before and after immobility, along each of the two degrees of freedom, in order to quantify the sequence in which they unfold. This procedure reveals that before immobility the maximal peak of translation ($T^*$) is exhibited before the maximal peak of rotation ($R^*$). After immobility, it is the maximal peak of rotation that is exhibited before the maximal peak of translation. To quantify the relative order of global peaks we calculate the difference between the time of maximal peak of rotation and the time of maximal peak of translation, namely, $t(R^*)-t(T^*)$. This procedure shows that translation precedes rotation before immobility, and that rotation precedes translation after immobility (see **Figure 6C**). While there is a very high variability in the time intervals across animals, all follow the same sequential order of global peaks.

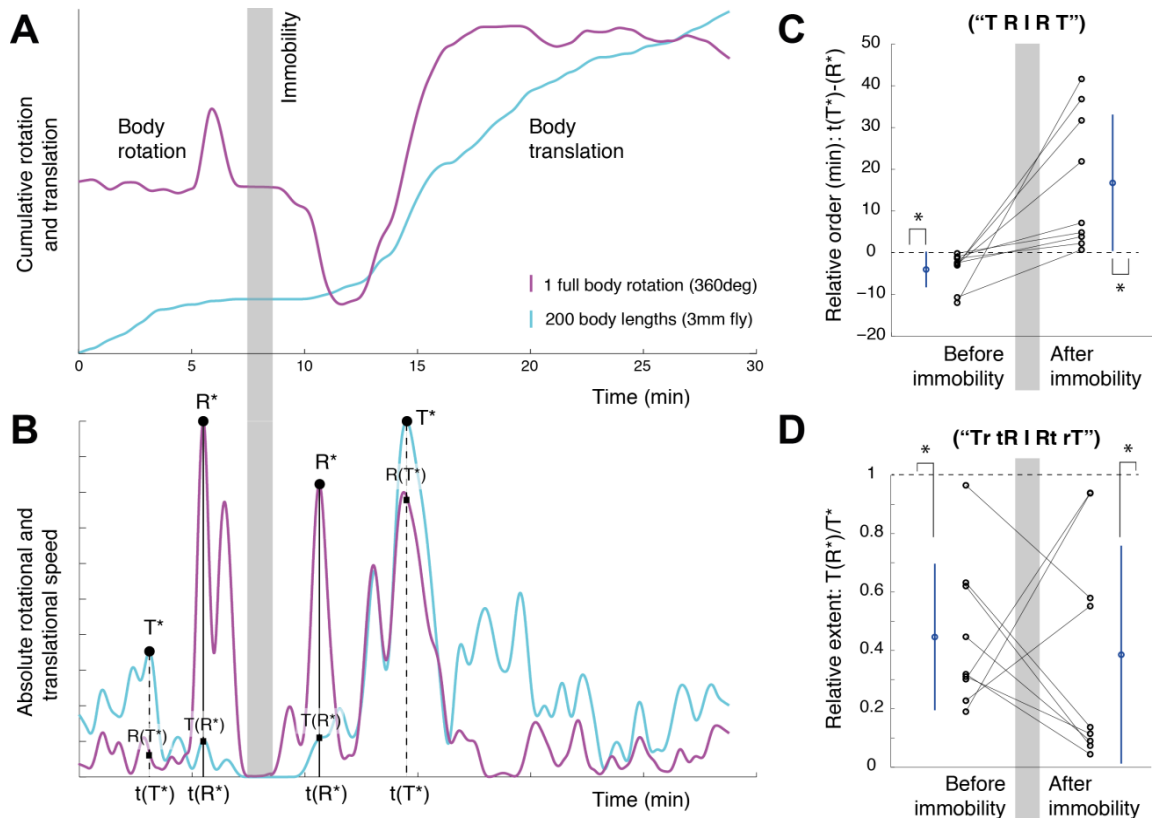

**Figure 6. Quantification of synchronic and diachronic dynamics of translation and rotation.** (**A**) Cumulative body rotation (magenta) and translation (cyan) reveals the global sequence of changes in the rotational and translational degrees of freedom. The shaded area marks the period of immobility which is used as a reference for measuring the events that precede and follow it. (**B**) Global changes in speed of progression and body orientation are obtained from the absolute time derivative of the curves in (A). As shown, a global peak in translation followed by a global peak in rotation precede immobility, and a global peak in rotation followed by a global peak in translation follow immobility (TRIRT). To characterize the strength of the reciprocal relations between global peaks we calculate the ratios between global peak of one degree of freedom, R(R*), and its value at the global peak of the other degree of freedom, R(T*). (**C**) Difference between the time of maximal peak of translation and the time of maximal peak of rotation, t(T*)-t(R*), is negative for transitions into immobility (p=0.004, Sign test) and positive for transitions out of immobility (p=0.004, Sign test). Absolute time differences are greater for transitions out of immobility than for transitions into immobility (p=0.0352, Sign test). Each line connecting dots represents the same animal. Mean and standard deviation in blue. (**D**) Quantification of the strength of reciprocity in the value of global peaks of rotation and translation during transitions in and out of immobility. The ratio T(R*)/T* is smaller than 1 for transitions in and out of immobility (p<0.004, Sign test). Dots represent the score for individual flies and lines connect the results for the same individual. The mean and standard deviation are colored in blue.

Next, in order to quantify the relative strength of the reciprocal relationship between global peaks, we calculate translation at its peak, T(T*), and compare it with translation when rotation is at its peak T(R*). Thus we measure the amount of reduction in translation by the time that rotation reaches its peak. The smaller the ratio T(R*)/T(T*), the stronger the phenomenon of reciprocity between translation and rotation (**Figure 6D**). Note that this relationship is by no means reciprocal at all times: in the graph presented there are cases where both rotation and translation increase, and also where both decrease together. In other words, rotation and translation are globally, *not* locally, reciprocal. Characterizing the relative strength of rotation and translation peaks by means of the above ratio is invariant to time rescaling and to absolute values of rotation and translation. This is necessary for capturing the invariance in the sequence and strength across individual animals, and particularly useful given the large variability in the timescales of unfolding of the phenomenon (some flies take minutes, others take hours) and in the rotation values (some flies perform ten full body rotations, others perform hundreds; and they do so at different rates).

On the whole, flies follow the same sequence of transition into immobility, involving, *for each dimension separately*, an enhancement, a reduction and then elimination of that degree of freedom, thus progressively narrowing down the fly's locomotor repertoire, and the same but opposite sequence of transition out of immobility, involving, *for each dimension separately*, an enhancement, and then subsiding to normal of that degree of freedom, thus progressively widening up the fly's locomotor repertoire. Such invariance can be summarized by the acronym TRIRT (Translation, Rotation, Immobility, Rotation, Translation).

***Behavior in and out of immobility: the rate of switching between directions of rotation.*** One predominant effect of cocaine is the high rate of repetition of full body rotations and their rotational speed. During rotation the animal may switch between clockwise (CW) and counterclockwise (CCW) rotations. It is now possible to examine the dynamics of switching in reference to immobility, in the context of the animal's freedom of movement. In asking this question, we focus on *how frequently* the fly changes the direction of rotation and *how biased* successive rotations are. Globally, there are long-term predominant biases to rotate in a particular direction. In particular, transitions out of immobility start by very long rotations in the same overall direction. However, the flies do not rotate monotonically in one direction but rather alternate between large amplitude rotations in one preferred direction and low amplitude rotations in the other direction (**Figure 7B-C**). Locally, the fly alternates between CW and CCW rotations. As shown in **Figure 7**, the switching rate decreases when going into immobility and increases when going out of immobility. As found for the synchronic relationship between translation and rotation (TRIRT), this diachronic pattern for rotational switching also exhibits mirror symmetry between the process leading to immobility and out of it (**Figure 7E-F**), and can be summarized by the acronim SsIsS (high Switching, reduced switching ability, Immobility, and the reverse).

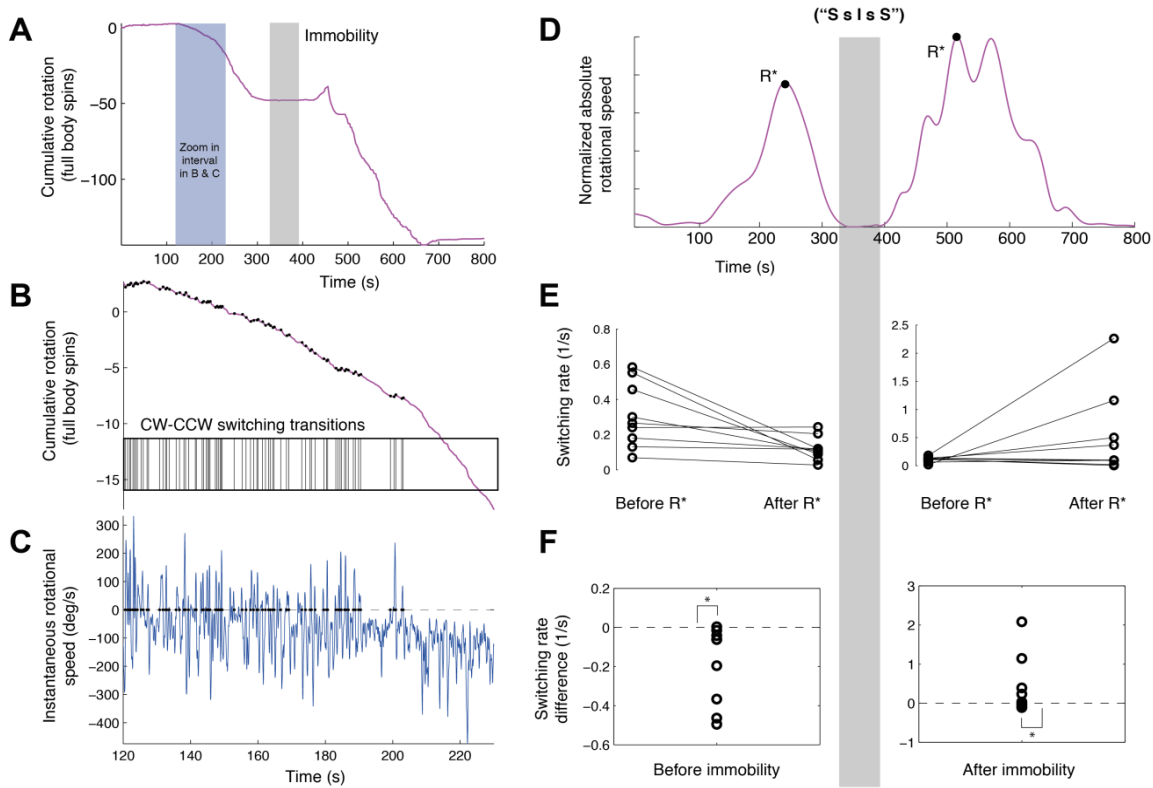

**Figure 7. Switching between clockwise and counterclockwise rotation decreases into immobility and increases out of immobility.** (**A**) Cumulative body orientation as a function of time for a single fly across the session. The grey shaded area marks the period of immobility. The blue shaded area marks a zoomed in time interval presented in (B) and (C). As shown, the general tendency of this fly is to rotate clockwise both before immobility and after immobility. Note that during transition into immobility the fly performs of the order of 50 full body rotations in 5 minutes, and during transition out of immobility, as much as 100 full body rotations in 5 minutes. (**B**) Zoomed in time segment in (A) of cumulative body rotation. As illustrated, while the fly generally rotates clockwise, it keeps switching between clockwise and counterclockwise rotations. Each transition is marked by a small black dot on the curve and a corresponding vertical bar on the transitions raster plot below. (**C**) Time derivative of the fly's momentary body orientation. Zero crossings in the rotational speed mark switching, corresponding to the vertical bars in (B). As shown, there is an overall drift in one direction, concurrent with a decrease in the number of transitions as a function of time. (**D**) Global changes in body orientation obtained from the absolute time derivative of the curve in (A). To examine the change in switching across the session, we partition the period leading to immobility into the segments preceding and following maximal rotation. We do the same for the period leading out of immobility. (**E**) Switching rate, calculated as the number of transitions per second, decreases in the interval preceding immobility (left panel) and increases in the interval following immobility (right panel). Dots represent the score for individual flies and lines connect the results for the same individual. (**F**) Switching rate decreases when going into immobility (left panel, p<0.039, Sign test) and increases when going out of immobility (right panel, p<0.039, Sign test). Dots represent the switching rate change for every animal.

## Discussion

***The generative rules that shape fruit fly cocaine induced locomotor behavior during the transition out and into immobility.***

*U*sing immobility as a reference for the measurement of behavior and cocaine as the parameter inducing a behavioral gradient we found that the flies exhibit a progressive expansion of their locomotor repertoire starting from immobility, and a progressive narrowing down of the repertoire leading into immobility. During expansion, for each key variable separately (fig. 6), the fly exhibits first enhancement and then reduction to normal values of movement along that variable: first, of body rotation in the horizontal plane, then of path curvature and then of speed of translation. The extents of movement across the key variables show reciprocal relations: when rotation is at its peak translation is low, and when translation is at its peak rotation is low, and path curvature is partly coupled to rotation. Transition into immobility from rich normal locomotor behavior unfolds by narrowing down of the repertoire in the opposite sequential order, also showing reciprocal relations between the extents of the same variables. Quantification of the generative rules of this behavior, based on the temporal sequence of global peaks of extent (fig. 6 C,D) and their reciprocal values provides a summary of the *bauplan* of this arthropod behavior, allowing a comparison to the rules reported in previous studies for movement into and out of immobility in vertebrates, where the behavior has been termed *The mobility gradient*[17].

*Vertebrates and fruit flies share the same generative rules*.
*Warmup*: In infant mice transition out of novelty-induced immobility consists of side-to-side head movements that increase in amplitude, gradually recruiting the forequarters, and then the hindquarters to extensive rotation in place around the hindquarters. Only after exhausting the horizontal plane by rotating around the hindquarters, forward stretching and subsequently forward translation appears, first along curved and then along straight paths ([movie S5](movie S5)). The same sequence is exhibited in amphibians[26], Fish[27], insectivores ([movie S6](movie S6)), and carnivores[28]. Head raising, forequarters raising and finally rearing on the hind legs are exhibited next[29,30]. The same sequence is exhibited both during moment-to-moment behavior and in ontogeny, during recovery from lateral hypothalamic damage[31]. Later in development, during, for example, play and ritualized fighting interactions, the inferior animal exhibits the less mobile portion of the sequence, culminating it with rearing and rotating around the hindquarters, whereas the superior may rear and rotate both around the hind legs and around the forelegs, exhibiting an expanded freedom of movement in both the horizontal and the vertical dimension[28,32]; for a review see[17,33] ([movie S7](movie S7)).
*Shutdown*: the opposite sequence, proceeding from rich normal behavior to enhanced, then reduced, then annulled movement, first of rearing, then of translation along straight, and then along curved paths, then of rotation, culminating in relative immobility is exhibited in rats following the administration of several dopamine agonists[4,5,34,35,36,37,38] ([Movie S8](Movie S8)). This mobility gradient[17,39], which is composed of warmup and shutdown sequences[29,30] shares with the mobility gradient exposed in fruit flies the same origin (immobility), knob (dopaminergic stimulation; but in vertebrates also novelty and proximity to a rival) and *bauplan*.

In the absence of fossils of behavior we temporarily suspend judgement about common descent. From the vantage point of comparative anatomy, *bauplans* are the core, or skeleton that resisted adaptation and around which there is a variable adaptive component. As shown by ethology, the core/adaptive distinction also applies to behavior[19,40].

***Searching for equivalent neurochemical substrates mediating the expansion and narrowing down of the animal's locomotor repertoire.***

Recently, it has been claimed that the vertebrate basal ganglia and the arthropod central complex are deeply homologous. In both, comparable systems of dopaminergic neurons, their projections, and dopaminergic receptor activities are involved in the modulation and maintenance of behavior[21]. Dopamine systems are also key players in generating and regulating the mobility gradient[23,24,34-38], and dopaminergic stimulation of specific substructures of the basal ganglia and central complex induce specific components of the mobility gradient respectively in rodents[41-45] and in flies[46]. The neurochemical processes mediating the vertebrate expansion of locomotor repertoire have been recently attributed by Cools and co-workers to dopaminergic feedforward loops operating in the basal ganglia[23,24]. The equivalence between the mobility gradient core phenomena and the feedforward loops exposed in the basal ganglia can be used as a search image or working hypothesis for studying the relations between the arthropod mobility gradient and the central complex. It might be useful to examine whether feedforward loops also mediate the expansion of locomotor behavior functions in the central complex.

**Methodological considerations.**

We found the following choices useful for uncovering common cross-phyletic generative rules of behavior

***In the study of behavior, we use the methodology practiced by comparative anatomy.*** Ethology's search for behavioral homologies was fulfilled only in closely related species. Founded by comparative anatomists, ethology aspired to provide a description of the fixed core of behavior, conceived to be resistant to past environmental pressures as well as to current experimental manipulations[40]. Two main discoveries that were claimed by classical ethology were that (i) behavior was an extension of brain anatomy, and therefore its structure must reflect the connectivity of the brain, and that (ii) homologies were as applicable to behavior as to anatomy[47,48]. The comparative study of behavior made a clear-cut distinction between structure and function, examining them from the point of view of a fixed core (e.g., grasping with foot in canaries) used in a variety of contexts (e.g., for nest building and for feeding[19]). Lorenz accepted his share of the Nobel Prize with the reflection that his "most important contribution to science" has been his discovery "that the very same methods of comparison, the same concepts of analogy and homology, are as applicable to characteristics of behavior as they are to those of morphology" [19]. In particular, to the same extent that anatomists use bones to establish the concept of skeletal homology and brain circuitries to refer to neural homologues, ethologists need to isolate the particulate processes of behavior in order to establish behavioral homologues.

*We prioritize the use of continuous kinematic variables over the use of ad hoc discrete patterns.*
The fundamental building block of classical ethology has been the so-called "behavior pattern", a species-specific coordination involving movements of some or all the parts of the animal's body, and performed one at a time by the animal[6,49,50]. These expert-established units were based, however, on form, which has not been part of the definition of anatomical homologies. For example, there is no similarity in the respective form and relative size of the bone identified as the radius in the forelimb of a human and of a whale, yet these bones are homologous. As already proposed in the 19th century, anatomical homology must be defined, not by its form, but by the relative positions and spatial relations between the elements of a structure, ("principle of connections" of St. Hillaire; "equivalence under transformation"[2,7,8,9]). The radii of the human and the whale thus derive their identity from the same relative position they occupy and their identical connectivity to the other bones of the respective forelimbs. Whenever we call a radius a radius we implicitly or explicitly imply a forelimb, which in turn implies a vertebrate's skeleton. In much the same way, the striatum of a bird and of a reptile, although very different in size and form, is identified as striatum because of its connectivity to other brain structures[51]. By parsing fly behavior prematurely into discrete patterns such as circling and locomotion[4], rather than representing them in terms of speed of translation, path curvature, and body orientation, the opportunity to derive the generative rules that constrain these variables is lost, and with it is lost the opportunity to discover the universality of the rules. Lorenz and followers studying duck display could not demonstrate homological sequences of patterns because the sequence was much too variable even across these closely related species, disallowing the claim that a pattern occupied the same relative position in two corresponding sequences performed by different species[3]. Ethology did succeed to expose behavioral homologies across closely related species (e.g.,[52,18]), but not across phyla.

*We describe the progressive unfolding of behavior from simple to complex and vice versa in reference to an origin presumably used by the organism itself*. As quantified in figures 5-7, demonstrating the invariant sequence could be accomplished by proceeding away from immobility in both directions so as to capture the serial emergence of first global peaks along each of the degrees of freedom in warm up and the serial performance of the last global peaks in shutdown. As degrees of freedom are added (or, respectively, eliminated) the sequence becomes increasingly unpredictable in warm up, or respectively stereotyped in shutdown, and the invariant sequence is lost. This applies both to the fly in this study and the mouse[30]. This is why sequential invariance (and hence homology) cannot be demonstrated in full blown behavior, which becomes increasingly differentiated, i.e., unpredictable, across performance.

*We prioritize structure over function.* The question whether homology should be based first and foremost on structure or on function engaged comparative anatomists for almost two hundred years[16]. While prioritizing structure has become common practice for almost a century with regard to brain and skeletal anatomy[13,14], behavioral neuroscientists have been fluctuating between prioritizing function over structure and vice versa to this day[15].

The "structure first" advocates[3,18,53] were aware of the need for data acquisition and analysis tools not available at the time, but missed the need for a descriptive technology highlighting the quantities that are actively managed by the organism (which was already available at the time[54]).

Current adherents of "function first", influenced by experimental psychology, Neo-Darwinism, and medical diagnosis, focus on the adaptive components of behavior, involving learning, habit formation, and normal locomotor performance, eg.,[21]. The descriptions they use are useful as long as the animal exhibits adaptive behavior, but fail with drug- or lesion- or genetically modified behavior. This is perhaps why the warm up and shutdown sequences (figure 1) were described as abnormal, aberrant, unusual and uncontrolled[6], and dopamine-manipulated behavior was described as ataxic, hypo- or hyperkinetic[21], missing the highly organized dynamics (figures 2-7). While a "function first" methodology thus considered warmup and shutdown as conserved[4] though aberrant[6] relics, a "structure first" methodology would treasure it as a crystalized manifestation of a cross phyletic invariance exposing the generative rules of a substantial component of arthropod and vertebrate locomotor behavior. In Goethe's words[55] "Organisms can transform their behavior into misshapen things not in defiance of law but in conformity with it, while at the same time, as if curbed with a bridle , they are forced to acknowledge its inevitable dominion". The dominion belongs to the generative rules.

***We prioritize generative rules over common descent explanations of behavior.*** The tension arising between these two ways of making sense of the diversity and unity of species that has been revealed through evolution has animated many biological debates for the last 180 years[2,17,26,56,57] and is currently experiencing a revived interest (https://royalsociety.org/events/2015/03/nervous-system/). The mainstream approach in neuroscience relies on historical explanations (common descent), embryonic lineages, and genetic programming[21]. A less common way, implemented in the current study, is based on demonstrating equivalence of structure by exposing the common cross phyletic generative rules that shape a diversity of forms through transformational unity. Controversy prevails, not only with regard to the primacy of structure but also with regard to how dependent is a structural definition of homology on the demonstration of a common descent. While Beer argues that "Clearly there has to be a way of judging homology independently of evolutionary considerations[3], and Wagner[13,14] separates homology from phylogeny by defining homology in terms of structure and development, Hodos and Atz maintain that behavioral homology must include a statement of the underlying structural elements that at least in principle can be traced to specific ancestral precursors[18,53]. The consistency of the phenotype (e.g., behavior) that species show despite much turnover in their gene pool[58]) is another argument for independent characterization of the phenotype. Characterization of homology on the basis of equivalence of generative rules is implemented by Goodwin[9] with regard to the morphogenesis of anatomical structures and by Golani[17] with regard to Kinematics.

Here we address the common architectures while suspending judgement about common descent, which, once demonstrated, will endow the *bauplan* with the status of a Darwinian homology. This procedure implements the heuristic potential of a map that sensitizes and guides the researcher to incipient, disguised, or missing parts of a structure[59,16] both at the neural and the behavioral levels.

***We apply the "serial homology" concept to situations where the same structure serves different functions in the same species.*** Using our *generative rules* as a search image, not only can we predict essential sameness in the brain/behavior interface of flies and rodents, but also anticipate that the same rules might underlie apparently different functional behaviors in the same species. This is similar to identifying serial homology in anatomy, where, e.g., hand and foot are considered

homologous because they share the same set of developmental constraints, caused by locally acting self-regulatory mechanisms of differentiation[14].

Reviewing the fruit fly larval behavior literature with a search image for low and high mobility one's attention is immediately captured by the abnormally high amounts of turning behavior exhibited by larvae with mutations in scribbler in the absence of food[60]. These appear to be respective manifestations of the high and low ends of the mobility gradient. The four key features characterizing low mobility in the cocaine treated fly – low speed of translation, highly curved path, high body rotation and immobility – exhibit a full correspondence to the features of the "abnormal crawling pattern" exhibited by scribbler larvae: low speed, curved paths, high turning rate, and long pauses[61]. The "knob" precipitating this behavior could be, as Sokolowski and co-workers suggest, the absence of food, or else, given our search image, the stress brought about by the absence of food, and even its presence in hungry flies[62,63]. Equivalent differences in mobility, expressed by pivoting and/or rearing on hind legs and forelegs, reported between vertebrate partners engaged in interactions[32,28], should be looked for in fruit fly courtship and agonistic interactions.

*Conclusion.* Homology is studied at all levels of biological organization from molecules to behavior[64]. Its nature is elusive [13,14], even more so with regard to behavior, where the distinction between the invariant conserved core and the adaptive component has been hardly described yet. There is a consensus with regard to the need for a comprehensive description of the equivalence of relations across all levels of the hierarchy all the way down to the molecular level and all the way back to common descent. But there is little awareness of the absence of cross phyletic homologues of behavior, the final common pathway of the hierarchy specifying the demand[17,65] on all the other levels. Homologues are indispensable because there cannot be a comparative science without having validated particulate processes that combine to exhibit the overall structure. Homology validates the partitioning of the whole into parts, based on the demonstrated equivalence under transformation. The state of the art with regard to the comparative study of baupläne were useful heuristic tools for searching for "missing" bones[59,16]. The same methodology can now be applied in phyletic and cross phyletic comparisons of behavior without reducing in any possible way the search for underlying equivalent connectivity at the neural, genetic and molecular levels, in the pursuit of common descent.

The generative rules derived from fruit fly cocaine-induced seemingly aberrant behavior are rules of differentiation and decay, exhibited on the three time scales of phylogeny[66], *ontogeny*[29,30], and *actual genesis*. Because they represent managed perceptual quantities they also define the organism's operational world, its *umwelt*[67]. In this way baupläne and *umwelten* are represented within the same phenomenological frame. Given the foundational role of behavior in neuroscience[68], it is perhaps time to start relating the various levels of the biological hierarchy to the behavioral *baupläne* they support, turning them -where possible- into Darwinian homologies.

## Materials and Methods

**Fly stocks.** Drosophila cultures were maintained at 24°C on a standard cornmeal-molasses medium in 12 hour light-dark cycle at 60% humidity. The experiments were performed on three-day-old flies of the wild-type laboratory strain Canton-S. Nine male flies were tested in a low-throughput high-content data approach.

**Behavioral arena.** The experimental setup for observing and tracking the flies was a 15 cm diameter circular arena with 0.7cm height wall and glass ceiling. The arena was illuminated from above with a 40W bulb. A camera placed above the arena recorded the fly's behavior. During the experiment there was a continuous airflow through the arena, through two small wall openings allowing also the introduction of the volatilized drug into the arena during the experiment[25].

**Animal preparation.** Neither food nor water was supplied to the fly during the entire experiment. All experiments were performed during the 12 hours light period, and on one fly at a time. Each fly was transferred to the arena and allowed to habituate there for one hour. Upon exposure to cocaine, the fly behavior was uninterruptedly recorded including full sedation (immobility) and the process of recovery (regaining normal locomotor behavior).

**Behavioral tracking.** The fly locomotor behavior was recorded at 25 frames per second using a CCD camera. Following video acquisition, the centroid position of the fly and its body axis direction were tracked with FTrack[69], a custom-made software written in Matlab (Mathworks). Raw trajectory data were corrected for tilt and rotation of the camera. Data segments during which it was not possible to assess the fly's orientation (fly located on the wall or jumping) were excluded from analysis.

**Behavioral analysis.** Quantitative analysis of the animal's behavior was based on the dynamics of three main degrees of freedom: centroid speed, centroid change in direction per unit of distance (path curvature) and body rotation (change in trunk orientation in the horizontal plane). Distinguishing between the direction of progression (centroid-based), the speed of progression, and the change in direction of rotations (body-based; where its longitudinal axis is facing) - the animal can, e.g., walk in one direction while facing with its rigid body any which way - north and then turn left, while still having the freedom to face north- allows to study the active management of where it is facing, where it is going, and how fast it is going[25,70,71]. Changes in the direction of progression are calculated per unit of progression and as a function of time in order to have a geometric curvature[72,73] in kinematic terms[74,75]. Switching between clockwise and counterclockwise body rotation was assessed via the zero-crossings of the instantaneous time derivative of the body angle, removing artifacts during arrests by pruning out rotations smaller than 12 degrees. In order to produce reliable estimates of local and global variables (as we study phenomena at different timescales: from sub-second small oscillations during rotation in place, to near-hour recovery trends) we use the variable-window smoothing LOWESS method[76]. Immobility, defined as the longest time interval of complete arrest (no translation nor rotation) across the whole session, allows to transform chronological time into activity, revealing dynamical invariants despite animal-to-animal behavioral variability.

**Acknowledgements.** This study was funded by the Portuguese Foundation for Science and Technology (FCT grant No SFRH/BPD/97544/2013 to AGM) and by the Israel Science Foundation (grant No 760/08 to IG and YB). We thank the participants of the "Homology in NeuroEthology" course held at the Champalimaud Neuroscience Programme for fruitful discussions. We acknowledge the gifts of fly stocks from Dani Segal. We thank Gonçalo Lopes, Eduardo Dias-Ferreira, Andre Brown, Lauren McElvain, Troy Shirangi, Eyal Gruntman and Ehud Fonio for valuable feedback on the manuscript.